\documentclass[lettersize,journal]{IEEEtran}
\pdfoutput=1
\usepackage{amsmath,amsfonts}
\usepackage{algorithmic}
\usepackage{algorithm}
\usepackage{array}
\usepackage[caption=false,font=normalsize,labelfont=sf,textfont=sf]{subfig}
\usepackage{textcomp}
\usepackage{stfloats}
\usepackage{url}
\usepackage{verbatim}
\usepackage{graphicx}
\usepackage{cite}
\usepackage{amsmath}
\usepackage{color, xcolor}
\usepackage{lineno}

\makeatletter
\let\NAT@parse\undefined
\makeatother
\usepackage[colorlinks, linkcolor=black, citecolor=black, anchorcolor=black,urlcolor=black]{hyperref}  

\begin{document}

\title{Seismic Data Strong Noise Attenuation Based on Diffusion Model and Principal Component Analysis}

\author{Junheng Peng, \IEEEmembership{Student Member, IEEE}, Yong Li, Zhangquan Liao, Xuben Wang, Xingyu Yang

\thanks{Junheng Peng is with the School of Geophysics, Chengdu University of Technology, Chengdu Sicuan, China}
\thanks{Yong Li is with the School of Geophysics, Chengdu University of Technology, the Key Laboratory of Earth Exploration \& Information Techniques of Ministry Education, and the College of Computer Science and Cyber Security, Chengdu University of Technology, Chengdu, Sichuan, China.}
\thanks{Zhangquan Liao is with the School of Geophysics, Chengdu University of Technology, Chengdu, Sichuan, China}
\thanks{Xuben Wang is with the School of Geophysics, Chengdu University of Technology, the Key Laboratory of Earth Exploration \& Information Techniques of Ministry Education, Chengdu University of Technology, Chengdu, Sichuan, China.}
\thanks{Xingyu Yang is with the Faculty of Geographical Science, Beijing Normal University, Beijing, China. }
\thanks{Corresponding author: Yong Li(e-mail: Liyong07@cdut.edu.cn)}}



\maketitle

\begin{abstract}
Seismic data noise processing is an important part of seismic exploration data processing, and the effect of noise elimination is directly related to the follow-up processing of data. In response to this problem, many authors have proposed methods based on rank reduction, sparse transformation, domain transformation, and deep learning. However, such methods are often not ideal when faced with strong noise. Therefore, we propose to use diffusion model theory for noise removal. The Bayesian equation is used to reverse the noise addition process, and the noise reduction work is divided into multiple steps to effectively deal with high-noise situations. Furthermore, we propose to evaluate the noise level of blind Gaussian seismic data using principal component analysis to determine the number of steps for noise reduction processing of seismic data. We train the model on synthetic data and validate it on field data through transfer learning. Experiments show that our proposed method can identify most of the noise with less signal leakage. This has positive significance for high-precision seismic exploration and future seismic data signal processing research.

\end{abstract}


\section{Introduction}
\IEEEPARstart As one of the main means of geophysical exploration, seismic exploration is often used in underground geological structure exploration and fossil resources exploration, which means that high-quality seismic data is indispensable for high-precision exploration work. However, the influence of various environmental conditions will lead to a certain disorder background noise in the seismic data, which has a great impact on the processing and interpretation of the seismic data, and then affects the subsequent work \cite{1}. Therefore, using a method to effectively remove the effects of this random noise is important for seismic data interpretation. At present, many researchers have proposed many methods to attenuate seismic noise, which can be mainly divided into two categories: conventional computing methods and deep learning methods. Among them, the conventional computing methods can be roughly divided into three categories: methods based on filters, methods based on sparse transforms, and methods based on rank reduction.

The first category includes filtering methods, where F-X predictive filtering was originally used to attenuate random noise in poststack seismic data and it assumes the predictability of linear seismic events in the frequency space (F-X) domain \cite{2} \cite{3} . In addition, many authors have proposed a variety of transform domain-based methods, which involve the second category of seismic noise attenuate methods, such as the Fourier transform \cite{4} \cite{9} \cite{10}, wavelet transform \cite{11} \cite{12}, seislet transform \cite{13}, and curvelet transform \cite{14}. They can convert seismic data into sparse domains for better separation of signal and noise. Because the correlation coefficient between two frequency componets equals the cosine of their phase-angle difference, Alsdorf \cite{5} proposed that this relationship could be used to build filters that separate noise from signals in seismic data in the F-X or F-K domains; Mousavi et al. \cite{6} proposed a method based on the synchrosqueezed continuous wavelet transform and custom thresholding of single-channel data to separate noise; Ahmed et al. \cite{7} first used seislet transform to estimate the signal components, and then apply the orthogonalization scheme to retrieve the leaked signal energy and restore it back to the initial signal estimate; Zu et al. \cite{8}  applied the curvelet transform to separate simultaneous sources based on the iterative soft-thresholding algorithm, which has also achieved good results in the separation of noise from seismic data.

The third category includes rank reduction. The methods based on rank reduction assume that the seismic data is a low-rank structure and the rank of the seismic data increases due to the addition of random noise, so the random noise can be attenuated by eliminating the increased rank of random noise \cite{15} \cite{16}. Many authors have improved this approach, and it has been extended to multichannel singular spectrum analysis which can process 3-D seismic data \cite{17}. In general, various conventional computing methods can handle some low level random noise attenuation of seismic data. However, conventional computing methods generally have a disadvantage that requires the selection of data-dependent parameters \cite{18}, and the selection of optimal parameters has become very difficult due to various factors in different exploration environments. Therefore, it is necessary to find an adaptive random noise attenuation method.

In recent years, with the continuous development of various fields of computers, deep learning (DL) methods have been widely used by many researchers \cite{19}. Among them, convolutional neural networks (CNN) are the most widely used DL backbone networks, and have achieved success in the fields of computer vision and natural language processing \cite{20} \cite{21}. On this basis, many authors have proposed the use of CNN-based methods for random noise attenuation. Yang et al. \cite{18} used a CNN with residual connections (Resnet) for the attenuation of random noise; Based on Resnet, Liao et al. \cite{22} proposed to use a feedback structure for multiple iterations to improve denoising performance by gradually reducing the input noise level and searching for the remaining signal from the estimated noise; Unlike these methods, Zhao et al. \cite{41} proposed to use denoising convolutional neural network (DnCNN) to directly learn noise extraction instead of noise attenuation, which is a blind gaussian denoising model and also works well to separate noise from seismic data. In addition, many authors proposed some improved methods based on denoising autoencoder (DAE) \cite{24} \cite{25} \cite{26} \cite{23}. In general, many improved methods based on DL have achieved certain results. However, for seismic data containing strong noise, the effect of these methods is often not ideal, and the robustness for data with different noise levels is not strong.

Inspired by the diffusion model \cite{42}, we argue that the noise attenuation process for seismic data can be transformed into a step-by-step prediction task through Bayesian equation. In the forward process, the noisy seismic data can be regarded as obtained by gradually adding noise to the clean seismic data according to a certain ratio. Therefore, the noise reduction process of seismic data can be regarded as the reverse of the forward process. We use the Bayesian equation to reverse the entire forward process, but there are still two problems: the first problem is that the random distribution items in the Bayesian equation cannot be obtained; and the second problem is the number of steps to restore cannot be obtained when processing blind Gaussian noise data. To solve the first problem, we train a DL model to predict the random noise added at each step of the forward process. And to solve the second problem, we refer to the work of Cheng et al. \cite{43} and propose to use principal component analysis (PCA) to analyze the noise level in seismic data and establish a method for determining the number of noise attenuation steps. Our experimental results show that it is necessary to perform noise estimation by PCA and obtain the number of noise removal steps, and the method we proposed can achieve better results than conventional methods when dealing with seismic data with strong noise, and can restore the subsurface structure information in seismic data.

We used an NVIDIA RTX2060 Laptop GPU with 6GB memory for the training and validation of experimental models, and through the experimental results combined with a variety of conventional computing methods, our proposed method achieved better results on seismic dataset with strong noise. The seismic dataset we used contains various geological structures. Compared with traditional methods, the method we propose can better preserve various geological structures, which is beneficial to the subsequent interpretation of seismic data.

\section{Proposed Methods}

\subsection{Diffusion Model}

The theory of noise diffusion is mainly divided into two parts, the forward process and the reverse process.

\subsubsection{Forward Process}
Let the real seismic data be $x_0$, and the noisy seismic data be $x_{t}$. We assume that the noise from $x_0$ to $x_t$ increases gradually, and the additional noise intensity in each time set to a sequence $\beta$, which can be seen in the Fig 1. Meanwhile, to minimize errors as much as possible, we set the $\beta$ values, which is the level of noise added at each step, to be very small, and the length of $\beta$ sequence is sufficiently long. So we generate a sequence in $x_0$ and $x_t$ with a length of t. The $m_{th}$ element in the sequence can be represented as:
\begin{figure}
\centerline{\includegraphics[width=8.274cm,height=2.13cm]{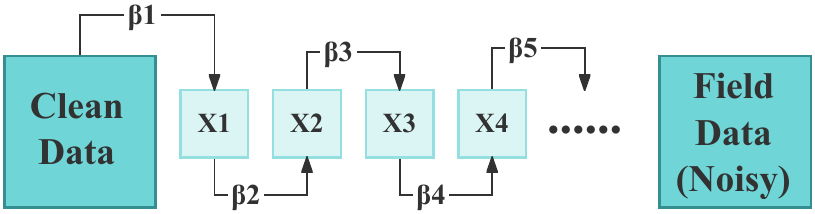}}
\caption{The addition of noise in the forward process.}
\end{figure}
\begin{equation}
x_m=\sqrt{\alpha_m}*x_{m-1} + \sqrt{1-\alpha_m}*z_m
\end{equation}
Among them:
\begin{equation}
\alpha_m=1-\beta_m
\end{equation}
$z_m$ represents Gaussian distributed noise at m moment. At the same time, the $(m-1)_{th}$ element can also be represented as:
\begin{equation}
x_{m-1}=\sqrt{\alpha_{m-1}}*x_{m-2} + \sqrt{1-\alpha_{m-1}}*z_{m-1}
\end{equation}
Bringing Equation 3 into Equation 1 yields:
\begin{equation}
\begin{split}
x_m=\sqrt{\alpha_{m-1}*\alpha_m}*x_{m-2} + \\
\sqrt{(1-\alpha_{m-1})*\alpha_m}*z_{m-1} + \sqrt{1-\alpha_m}*z_m
\end{split}
\end{equation}
where $z_m$ and $z_{m-1}$ are both Gaussian distributed noise. Because the sum of two independent Gaussian distributions is also a Gaussian distribution, we use $z^`$ to represent both of them, which can be expressed as:
\begin{equation}
\begin{split}
\sqrt{(1-\alpha_{m-1})*\alpha_m}*z_{m-1} + \sqrt{1-\alpha_m}*z_m \\
= \sqrt{1-\alpha_{m-1}*\alpha_m}*z^`
\end{split}
\end{equation}
Thus, $x_m$ can be represented as:
\begin{equation}
x_m=\sqrt{\alpha_{m-1}*\alpha_m}*x_{m-2} + \sqrt{1-\alpha_{m-1}*\alpha_m}*z^`
\end{equation}
In the same way, multiple substitutions can yield:
\begin{equation}
x_m=\sqrt{\overline{\alpha}_m}*x_{0} + \sqrt{1-\overline{\alpha}_m}*\hat{z}_m
\end{equation}
where $\overline{\alpha}_m$ represents the cumulative multiplication of $\alpha_1$ to $\alpha_m$ and $\hat{z}_m$ represents the combination of $z_1$ to $z_m$ Gaussian distributions. Therefore, the noisy seismic data $x_t$ can be expressed as:
\begin{equation}
x_t=\sqrt{\overline{\alpha}_t}*x_{0} + \sqrt{1-\overline{\alpha}_t}*\hat{z}_t
\end{equation}
Therefore, in the whole forward process, the formula for directly calculating the noisy data $x_t$ from the clean data $x_0$ is derived.

\subsubsection{Reverse Process}
In the process of reverse, we reversely solve $x_{t-1}$ through the Bayesian equation, which can be expressed as:
\begin{equation}
P(x_{t-1}|x_t, x_0)=P(x_t|x_{t-1}, x_0)*\frac{P(x_{t-1},x_0)}{P(x_t,x_0)}
\end{equation}
where the right-hand side of the equation can be expressed as:
\begin{equation}
P(x_t|x_{t-1}, x_0)=\sqrt{\alpha_{t}}*x_{t-1} + \sqrt{1-\alpha_{t}}*z_{t}
\end{equation}
\begin{equation}
P(x_{t-1},x_0)=\sqrt{\overline{\alpha}_{t-1}}*x_{0} + \sqrt{1-\overline{\alpha}_{t-1}}*\hat{z}_{t-1}
\end{equation}
\begin{equation}
P(x_t,x_0)=\sqrt{\overline{\alpha}_{t}}*x_{0} + \sqrt{1-\overline{\alpha}_{t}}*\hat{z}_{t}
\end{equation}

According to other related work \cite{42} \cite{44} \cite{45}, we expand them using Gaussian distribution function:
\begin{equation}
x\sim N(\mu, \sigma^2)
\end{equation}
\begin{equation}
f(x)=\frac{1}{\sqrt{2*\pi}*\sigma}*exp(-\frac{(x-\mu)^2}{2*\sigma^2})
\end{equation}

Expand Equation 9, we can obtain:
\begin{equation}
x_{t-1}\sim N(\mu_{t-1}, \sigma_{t-1}^2)
\end{equation}
\begin{equation}
\mu_{t-1} = \frac{1}{\sqrt{\alpha_t}}*(x_t - \frac{1-\alpha_t}{\sqrt{1-\overline{\alpha}_t}}*\hat{z_t})
\end{equation}
\begin{equation}
\sigma_{t-1}^2=\frac{(1-\overline{\alpha}_{t-1})*\beta_t}{1-\overline{\alpha}_{t}}
\end{equation}

So far, throughout the reverse process, we have used $x_t$ to deduce $x_{t-1}$. At the same time, this theory has been proved to be effective by many works \cite{46} \cite{47} \cite{48}, using $x_t$ to solve $x_0$ step by step. However, there are also two problems here.

The first problems is how to predict the noise $\hat{z_t}$ added in each step. Like many works mentioned before, we use DL models for prediction. We will introduce the network structure used in this work in detail later.

The second problem is how to determine the $t$ of the data to be processed, that is, how to determine the number of times the data needs to be reversed.

\subsection{Principal Component Analysis}

In the process of field exploration, the noise level of the obtained seismic data is often unavailable. Therefore, after referring to similar works \cite{43} of Chen et al., we choose to use PCA to estimate the noise level, and establish a statistical relationship equation with $t$. In this work, after extensive experiments, we choose the $\beta$ sequence used in our work satisfies:
\begin{equation}
\beta_t = 0.00115 + 0.00015 *t
\end{equation}
where $t$ ranges from 1 to 200. Then, PCA noise analysis is performed on the data at each $t$ to establish a statistical relationship.

Chen et al. first divided the data into multiple patches, and turned two-dimensional or three-dimensional patches into one-dimensional vectors. Then perform PCA operation on each patch, find the eigenvalues and sort them from large to small. On this basis, Chen et al. divided the eigenvalues into two parts: the eigenvalues of the main dimension and the eigenvalues of the redundant dimension. In the redundant dimension, that is, the noise dimension, its eigenvalues satisfy the Gaussian distribution, which mean is equal to its variance.

According to the follow-up work of Chen et al., when there are enough redundant dimensions, it can be judged whether there is a main dimension by judging whether the average value and median value of the eigenvalues of all dimensions are equal. By continuously eliminating the main dimension, the redundant dimension is finally obtained, and the standard deviation of its eigenvalue is the noise level. In the Fig 2 are examples of the eigenvalue distribution of several patches.
\begin{figure}
\centerline{\includegraphics[width=8.495cm,height=5.65cm]{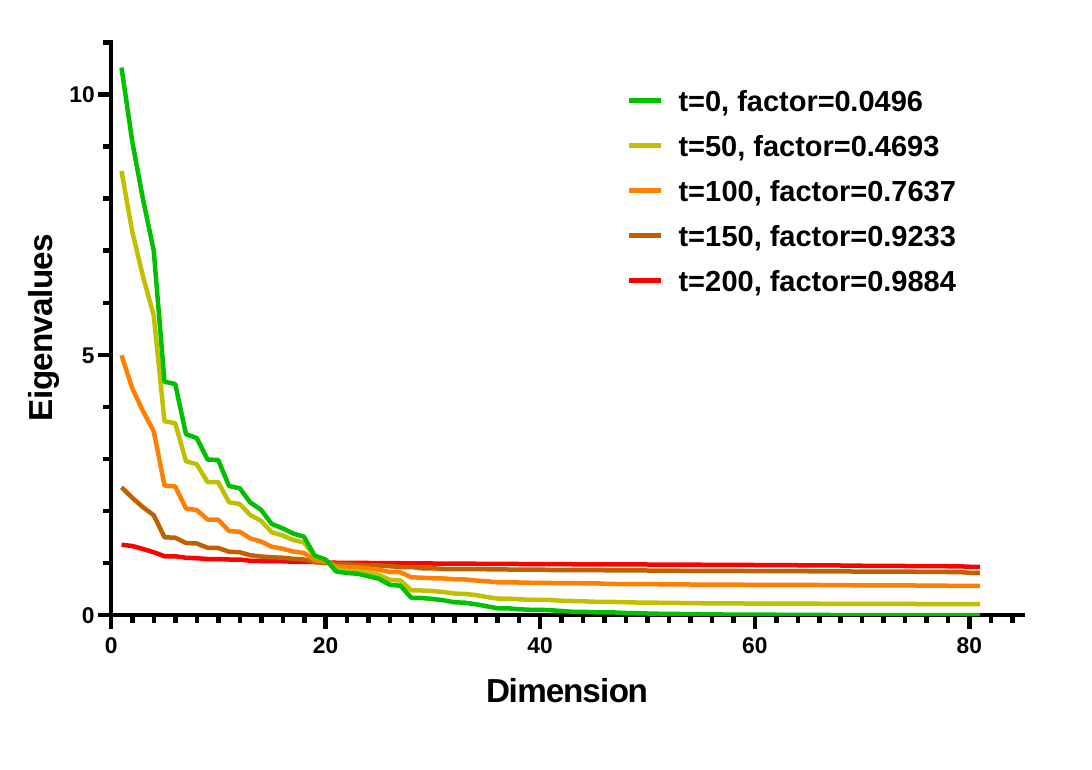}}
\caption{Some examples of the eigenvalue distribution.}
\end{figure}

Based on this work, we computed the average noise level and signal-to-noise ratio (SNR) for multiple sets of seismic data with different $t$. Its relationship curve is shown in the Fig 3. To quantify their relationship, we fit it using third order polynomial, which can be expressed as:
\begin{figure}
\centerline{\includegraphics[width=9.656cm,height=4.572cm]{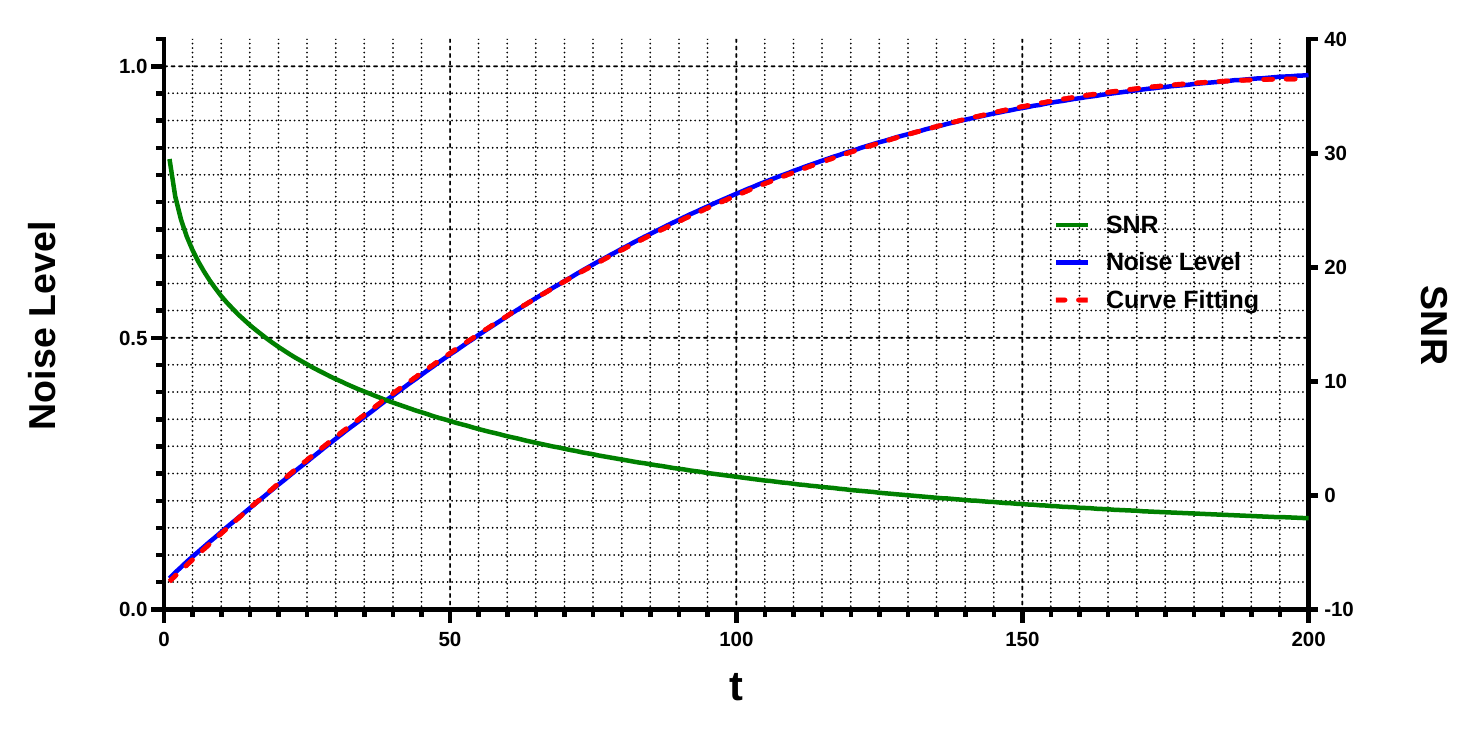}}
\caption{The curves of SNR and noise level.}
\end{figure}
\begin{equation}
norm\ factor = B_0+B_1*t+B_2*t^2+B_3*t^3
\end{equation}
because in our theory, $t$ can only take an integer between 1 and 200. Therefore, choose an appropriate $t$ to make the calculated $norm\ factor$ the closest to the real factor value. This $t$ is the number of steps that need to be reversed for the data. After fitting according to a large amount of data, we can get:
\begin{equation}
B_0 = 0.04170
\end{equation}
\begin{equation}
B_1 = 0.01009
\end{equation}
\begin{equation}
B_2 = -3.066 * 10^{-5}
\end{equation}
\begin{equation}
B_3 = 1.797 * 10^{-8}
\end{equation}

At the same time, we use R-squared to evaluate the goodness of fit, which can be expressed as:
\begin{equation}
R^2 = 1-\frac{\sum_{i=1}^{n}(\hat{y_i}-y_i)^2}{\sum_{i=1}^{n}(\overline{y}-y_i)^2}
\end{equation}
where $y$ represents the true value, $\hat{y}$ represents the fitting result, and $\overline{y}$ represents the average value of the true value. R-squared is often used in various regression tasks to detect the goodness of fit, and in our work, the $R^2$ of the fitting equation is 0.9998, which can be seen as a good fit.

The above $norm\ factor$ is obtained on some data sets, and the data values in the data sets are in a stable range. At the same time, the overall value range of the data has a positive correlation with the noise level estimate factor. Therefore, when using other data, we need to adjust the data as a whole. This adjustment can be expressed as:
\begin{equation}
norm\_X = \frac{X*A}{\frac{1}{n}*\sum_{i=1}^{n}{|x_i|}}
\end{equation}
Among them, $X$ represents the seismic data, and $A$ is the absolute value mean calculated by us in the data set from which the $norm\ factor$ function is obtained. In our field data experiments, it is shown that using $norm\_X$ instead of $X$ to calculate $t$, the calculated $t$ is closer to the actual number of steps that need to be restored.
\begin{equation}
A=0.7917
\end{equation}

To sum up, in this section, we establish a system for evaluating the seismic noise level of field seismic data and calculating the number of steps it needs to be reversed. The whole calculation process of $t$ is shown in Algorithm 1.
\begin{algorithm}[!h]
\caption{Calculate $t$}
\begin{algorithmic}[1]
\STATE $norm\_X = (X*0.7917)/(\frac{1}{n}*\sum_{i=1}^{n}{|x_i|})$
\STATE $norm\ factor=NoiseEstimate(norm\_X)$
\FOR{$t=1$ to $200$}
\IF{$(0.04170+0.01009*t-3.066*10^{-5}*t^2+1.797*10^{-8}*t^3)\approx norm\ factor$}
\RETURN{$t$}
\ENDIF
\ENDFOR
\end{algorithmic}
\end{algorithm}

\subsection{Model Structure and Training}

\subsubsection{Model Structure}

In this work, we used a network improved based on Unet network structure, and the network structure we used is shown in the Fig 4. Like other work \cite{42} \cite{44}, we use Attention, ResnetBlock, LinearAttention as the backbone of the network. The data input to the network includes $x_t$ and the $t$ corresponding to $x_t$. When $x_t$ is gradually processed, down-sampled and up-sampled, we use sinusoidal positional embeddings (SPE) \cite{49} to encode $t$ and $x_t$, while SPE can ensure that the position embeddings is bounded.
\begin{figure}
\centerline{\includegraphics[width=8.839cm,height=4.8722cm]{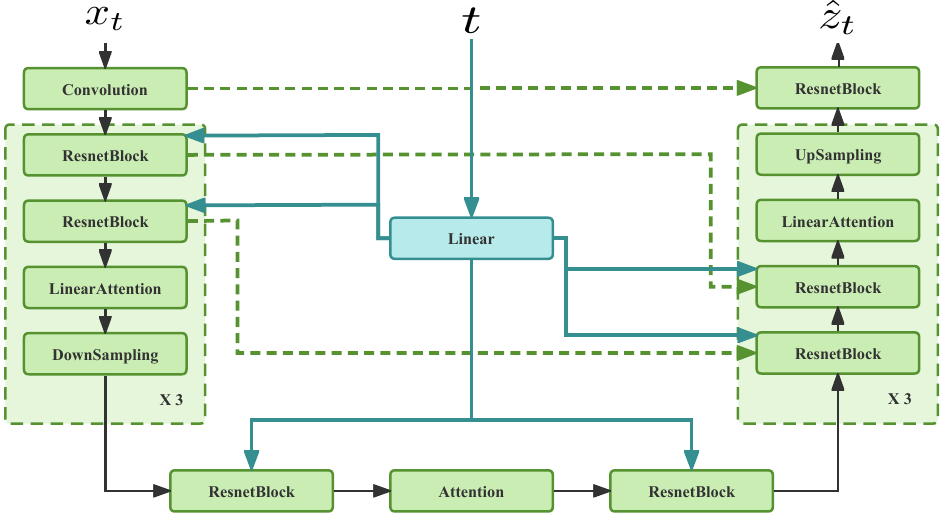}}
\caption{The structure of model.}
\end{figure}

The model has been uploaded to our github repository \cite{50}, including the settings of various parameters and model details. We use this network in our work to predict $\hat{z}_t$ through $x_t$ and $t$ as much as possible, and gradually eliminate the noise in seismic data through the formula derived above.

\subsubsection{Training Strategy}

In our work, in order to enable the proposed model to learn the characteristics of $\hat{z}_t$, we use the Adam optimizer and the $L1$ $Loss$ function to update the gradient of the parameters. The $L1$ $Loss$ function can be expressed as:
\begin{equation}
L1\ Loss = \frac{1}{n}*\sum_{i=1}^{n}|(\hat{z}_t-Model(x_t, t))|^2
\end{equation}

The training process of the model is described in Algorithm 2, and the verification of the model is described in Algorithm 3.
\begin{algorithm}[!h]
\caption{Training}
\begin{algorithmic}[1]
\FOR{$i=1$ to $Max\ Iteration$}
\STATE $x_0\sim Dataset(x_0)$
\STATE $t\sim Uniform(\{1, ..., 200\})$
\STATE $\hat{z}_t\sim N(0, I)$
\STATE $x_t=\sqrt{\overline{\alpha}_t}*x_{0} + \sqrt{1-\overline{\alpha}_t}*\hat{z}_t$
\STATE $Adam(Model, \frac{1}{n}*\sum_{i=1}^{n}|(\hat{z}_t-Model(x_t, t))|^2)$
\ENDFOR
\end{algorithmic}
\end{algorithm}

\begin{algorithm}[!h]
\caption{Validation}
\begin{algorithmic}[1]
\STATE $Input\ Field\ Noisy\ Data:\ x$
\STATE $T = NoisyAnalysis(x)$
\STATE $a=\frac{1}{n}*\sum_{i=1}^{n}{|x_i|}$
\STATE $x_T=x*0.7917\div a$
\FOR{$t=T,...,1$}
\STATE $z\sim N(0,I)\ if\ t>1, else\ z=0$
\STATE $\hat{z}_t = Model(x_t,t)$
\STATE $x_{t-1}=\frac{1}{\sqrt{\alpha_t}}*(x_t-\frac{1-\alpha_t}{\sqrt{1-\overline{\alpha}_t}}*\hat{z}_t)+\sigma_t*z$
\ENDFOR
\STATE $x_0=x_0*a\div 0.7917$
\RETURN{$x_0$}
\end{algorithmic}
\end{algorithm}

We use NVIDIA RTX2060 Laptop GPU with 6GB memory for model training and verification. The training took about 8 hours in total, and the loss decline curve is shown in the Fig 5. More comparative experiments will be shown in the next section.
\begin{figure}
\centerline{\includegraphics[width=8.5cm,height=5.5cm]{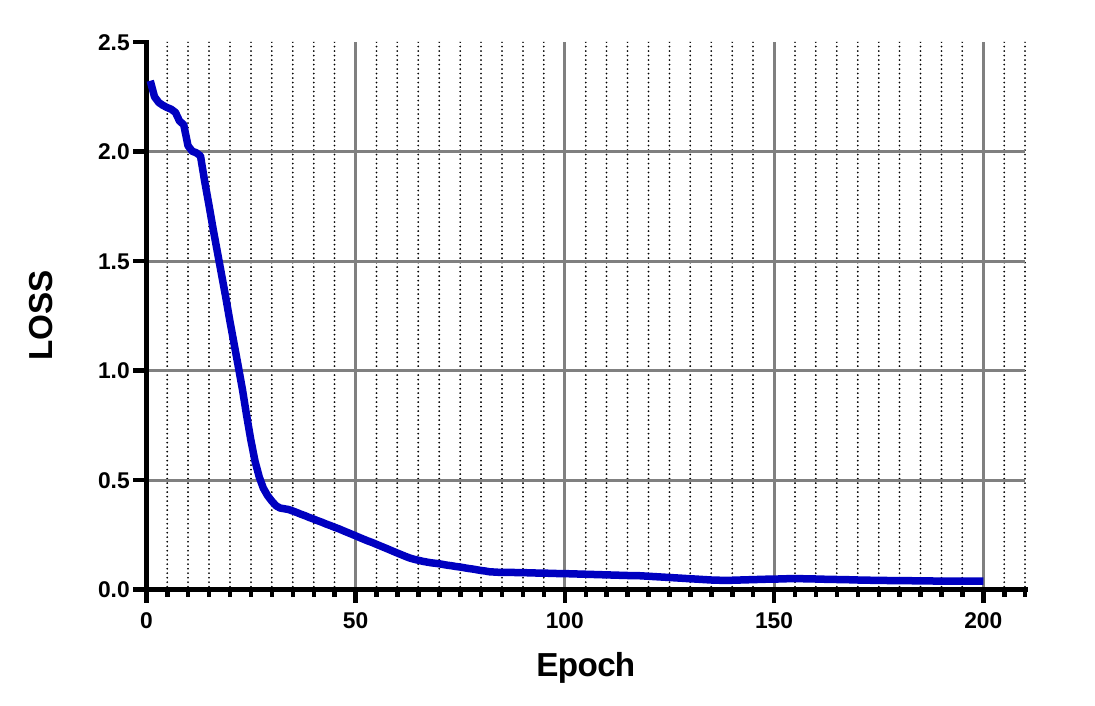}}
\caption{The loss decline curve.}
\end{figure}

\section{Experiment}

\subsection{Methods for Comparison}

In this experiment, the first comparison method we used is f-x deconvolution. F-x deconvolution (f-x decon) is a seismic data denoising method based on the assumption that seismic data with a linear coaxial axis can be autoregressively represented for each frequency slice in the f-x domain \cite{36}. F-x decon is a relatively mature method for noise attenuation in seismic data, so in our experiment, we first choose f-x decon to process noisy seismic data. We first split the data into multiple 40*40 patches, each with 80$\%$ overlap with neighboring patches. These patches are f-x decon processed and then restored to seismic data.

The second comparison method we used is hankel sparse low-rank approximation (HSLR) \cite{54}. Anvari et al. \cite{54} proposed a multivariate generalization of the minimax-concave penalty (MCP) function inducing sparsity on seismic data in the time-space domain. They decomposed sparse representation of data into semi low-rank, and defined the sparse components with the best approximate of noisy measurement matrix, by extracting low-rank matrix using optimal (re)weighting of the singular vectors of the observed matrix. HSLR has achieved very good results on various data, so we choose to use HSLR to compare and evaluate our proposed method.

The third comparison method we used is a twice denoising autoencoder framework (TDAE) \cite{55}. On the basis of DAE, liao et al. \cite{55} added a data generator. The DAE attenuates random noise without ground truth and works with a vectoring patching technique to reduce time complexity. In the data generator, local correlation (LC) is first developed to nonlinear LC to detect and extract the signal leakage with noise components suppressed. Then, the extracted signal leakage is compensated back to the DAE output in a supersaturated way to generate a new record. After that, DAE is used again to suppress the remaining noise in the new record. It has been shown in experiments that TDAE has a good effect on random noise attenuation of field data.

In our experiments, we compare these three methods to evaluate the performance of our proposed model.

\subsection{Description of Dataset}

In our experiment this time, we used 1000 synthetic data for model training, which included geological structures involved in various actual exploration tasks \cite{60}. After training, we first use 100 additional synthetic data for performance evaluation. Then we process the seismic data in the field and make a comparative evaluation through the migration of model parameters. The Fig 6 shows some examples of the synthetic data we use.
\begin{figure}
\centerline{\includegraphics[width=8.871cm,height=10.64cm]{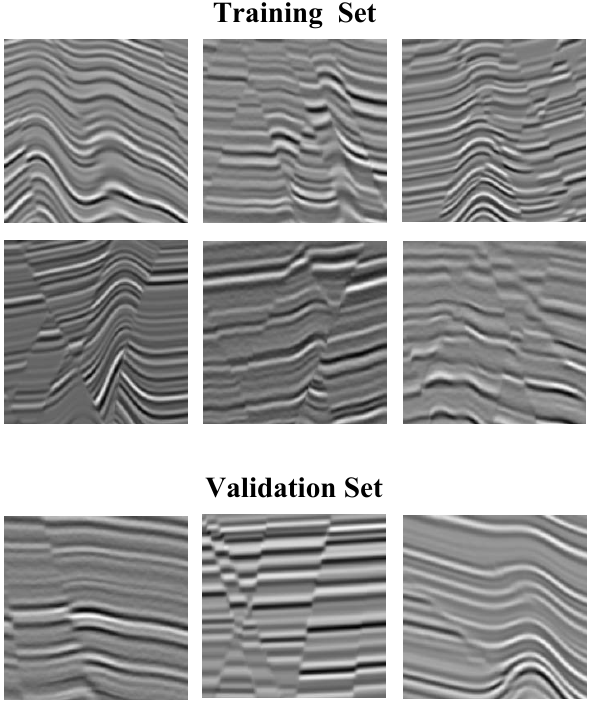}}
\caption{Some examples of the dataset.}
\end{figure}

In addition to synthetic data, we use two field data sets, $Kerry-3D$ and $F3\ Netherlands$, to verify the performance of our proposed method. They are two publicly available datasets that can be found in the related websites \cite{52} \cite{53}. Among them, $Kerry-3D$ is usually used in the verification work of fault detection tasks, which contains various degrees of noise; $F3\ Netherlands$ is very similar to $Kerry-3D$, and also contains different degrees of noise, and is usually used as a verification experiment for various noise attenuation tasks. In our work, we selected some patches with strong noise in these two data for verification experiment.

\subsection{Experiment on Synthetic Data}

\begin{figure}
\centerline{\includegraphics[width=8.8375cm,height=3.325cm]{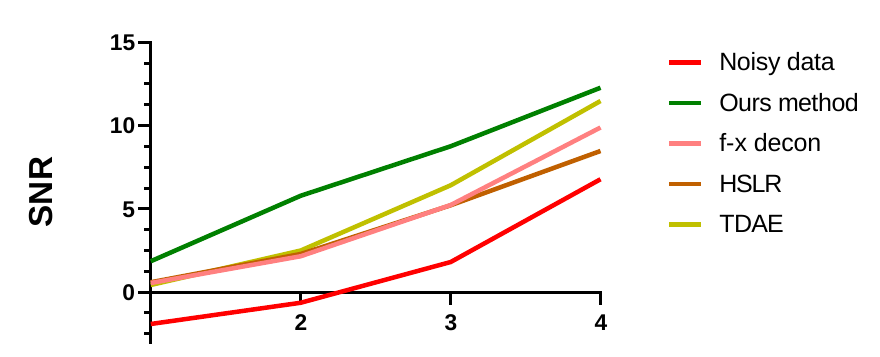}}
\caption{Average SNR of the denoised data.}
\end{figure}

In this part of experiment, we first add noise to synthetic validation data set to generate noisy data. We generated four levels of random noise in total: $t=50$, $t=100$, $t=150$, $t=200$. On this dataset, the average SNR of our proposed method and competing methods is shown in the Fig 7. It can be seen from the figure that compared with other competing methods, our proposed method has different degrees of improvement in both data with high noise level and low noise level. In addition, we selected several examples for visualization, as shown in Fig 8. The examples we have chosen include folds and faults that are two common geological formations.

\begin{figure}
\centerline{\includegraphics[width=8.215cm,height=15.07cm]{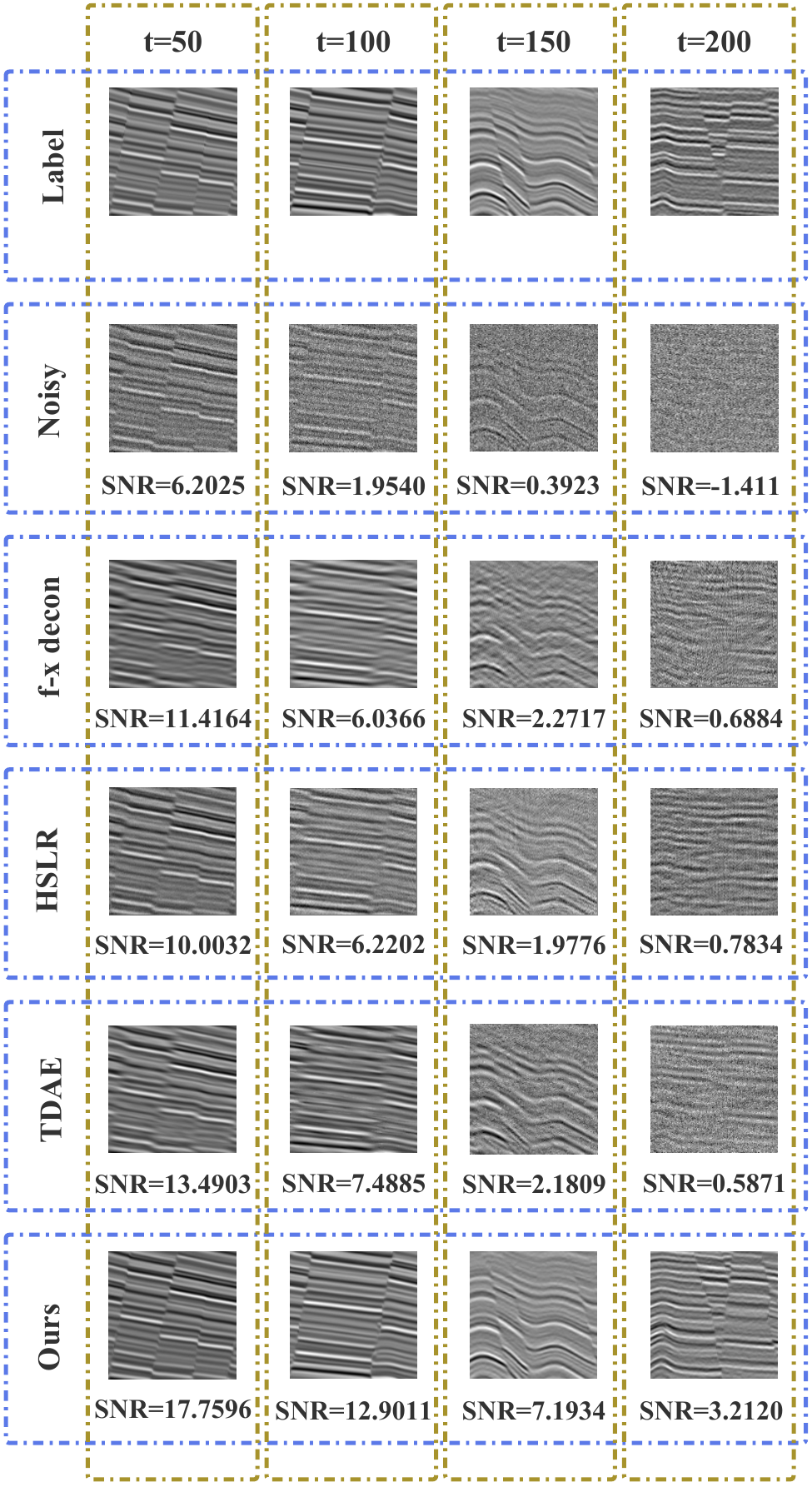}}
\caption{Some examples of the denoised data.}
\end{figure}

In the comparison between $t=50$ and $t=100$, f-x decon, HSLR and TDAE can barely eliminate the noise in the seismic data and restore the fault structure; while our proposed method restores data to the level of the label as much as possible, the fault boundaries are clearer. In the comparison of $t=150$, the noise at this time can be regarded as strong noise. The effects of f-x decon, HSLR and TDAE have all declined, and some data can be restored, but most of the data are blurred or even wrong; at this time, the method we proposed still maintains a certain effect, and can clearly restore faults and folds , most of the noise is eliminated. Finally, we conducted a comparative experiment at $t=200$, and the proportion of effective signals is extremely low at this time. It can be seen that the shape of the effective signal is hardly observed at this time. Conventional methods such as F-x decon, HSLR, and TDAE are all invalid at this time, and there are a lot of errors in the processed results. In contrast, our proposed method can still restore most of the effective information with very few effective signals in the case of extremely strong noise, although there are a small number of errors. As can be seen from the figure, our proposed method can restore the characteristics of folds and faults, while other competing methods have failed.

In addition, we computed residual maps for each result, which can be expressed as:
\begin{equation}
Residual\ Map = Label - Result
\end{equation}
as shown in the Fig 9, our proposed method has the least residual error compared with other methods, which means its ability of noise removal is stronger.
\begin{figure}
\centerline{\includegraphics[width=8.8cm,height=7.975cm]{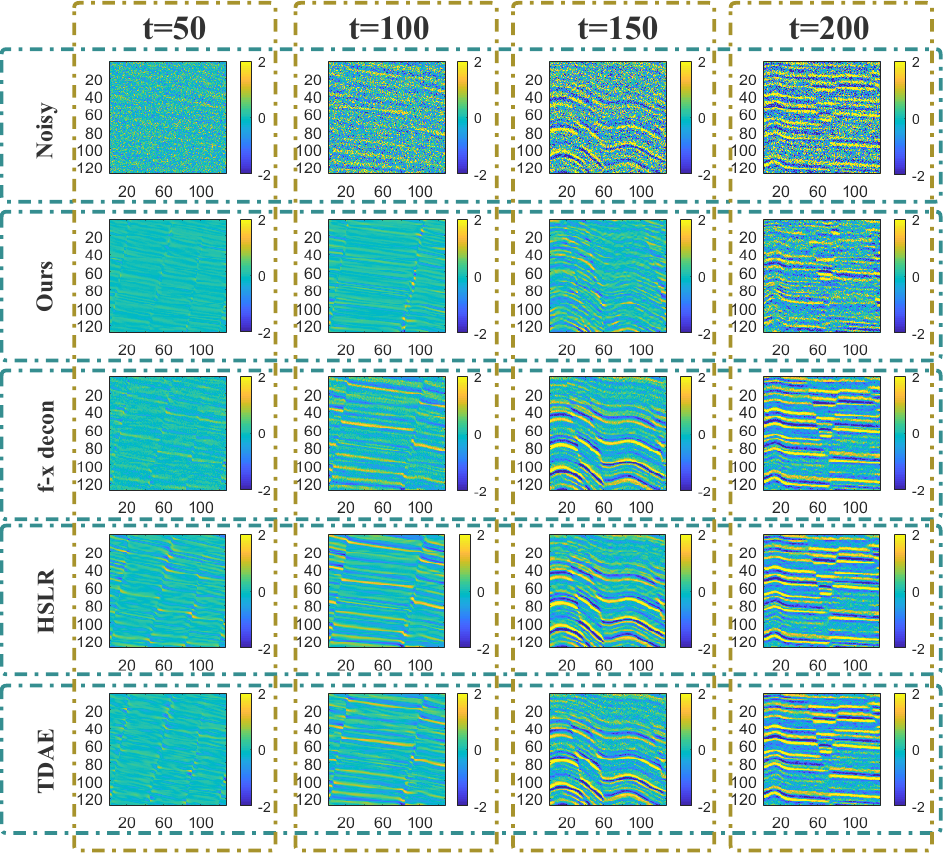}}
\caption{Some examples of the denoised data.}
\end{figure}

In order to further analyze the effect of our proposed method, we selected the data at $t=200$ to show the noise removal process of our proposed method, which as shown in the Fig 10. We took a data that requires 200 steps of noise removal processing as an example. It can be seen that although the noise level of the input data is extremely high, our proposed method can restore certain stratigraphic data after about 75 steps of noise removal processing, and then gradually restore the details of the data on this basis.

\begin{figure}
\centerline{\includegraphics[width=8.648cm,height=10.064cm]{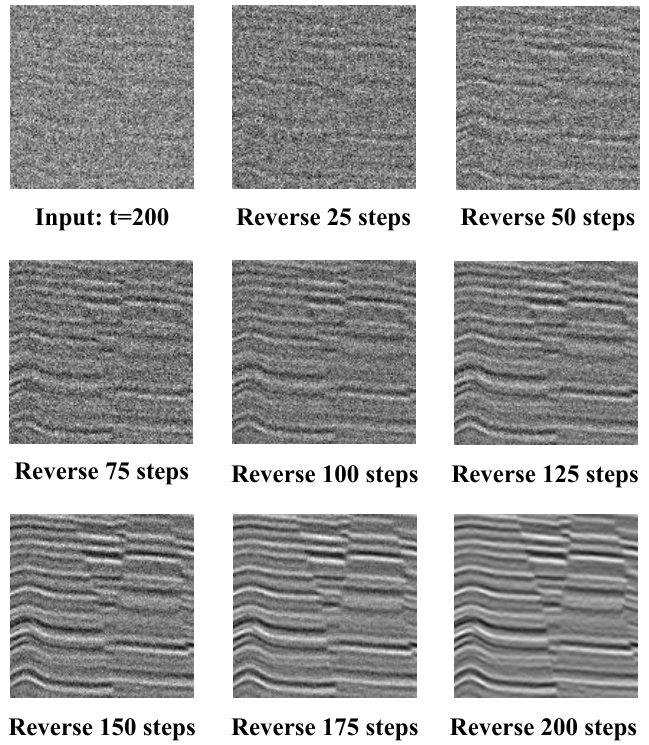}}
\caption{The entire noise removal process for an example. }
\end{figure}

Comparative experiments on synthetic data show that our proposed method outperforms other competing methods in the face of extremely strong noise.

\subsection{Experiment about Noise Level Estimation}

In our work, we propose to use methods such as PCA to determine the number of steps $t$ that the data needs to be restored. If only the diffusion model \cite{42} is used and $t$ is artificially determined, there will be a problem of excessive or insufficient $t$. We conduct a set of experiments to demonstrate the importance of estimating the noise level. Our experimental results are shown in the Fig 11.
\begin{figure}
\centerline{\includegraphics[width=8.356cm,height=4.132cm]{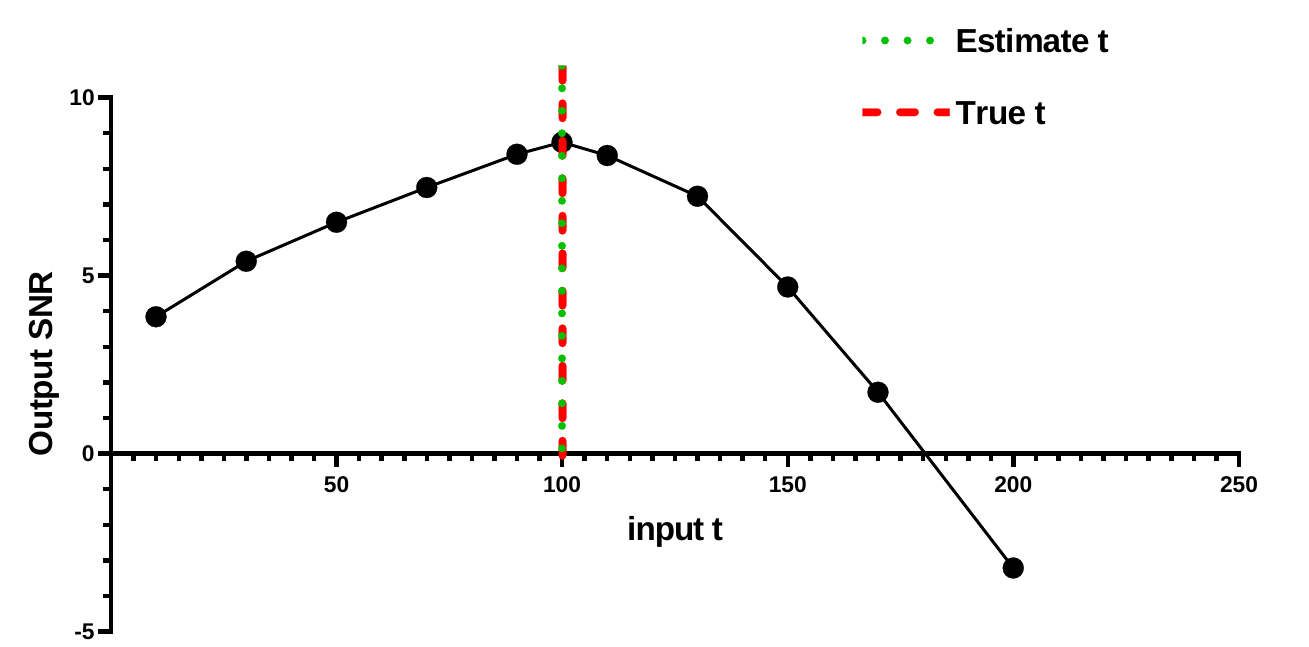}}
\caption{The SNR of the results corresponding to different inputs $t$. }
\end{figure}

It can be clearly seen that if we do not use PCA to evaluate the noise level of seismic data to obtain $t$ and directly set it artificially, no matter whether the input $t$ is too large or too small, the final effect will be reduced. However, the noise estimation method we propose can estimate the real $t$ well, thus effectively improving the effect of the diffusion model. Therefore, we believe that it is necessary to use our proposed method to estimate $t$ of the data when performing blind Gaussian noise processing.

\subsection{Experiment on Field Data}

\subsubsection{Transfer Learning}

Transfer learning is a commonly used method in the field of DL and has been adopted in many existing works \cite{56} \cite{57} \cite{58}. In our work this time, we use synthetic data for model training, and transfer the trained parameters for field data verification. Transfer learning effectively alleviates the shortage of field-labeled data, avoids repeated training of the model, and saves computing resources. Experiments show that our proposed method can achieve good performance on field data even through transfer learning.

\subsubsection{Experiment on F3 Netherlands}

In this section of the experiment, we selected the patch with high noise in the $F3\ Netherlands$ dataset for the experiment, and the experimental results are shown in the Fig 12. It can be seen that our proposed method can identify more noise than other methods, and the processed results are more effective in details. In order to analyze the effect difference more intuitively, we quantitatively analyze the noise removal effects of the four methods using the average local similarity \cite{59}. The average local similarity is obtained by calculating the average value of the local similarity, as shown in Table 1. It can be seen that our proposed method can not only identify most of the noise, but also have the least signal leakage.
\begin{figure}
\centerline{\includegraphics[width=8.833cm,height=15.697cm]{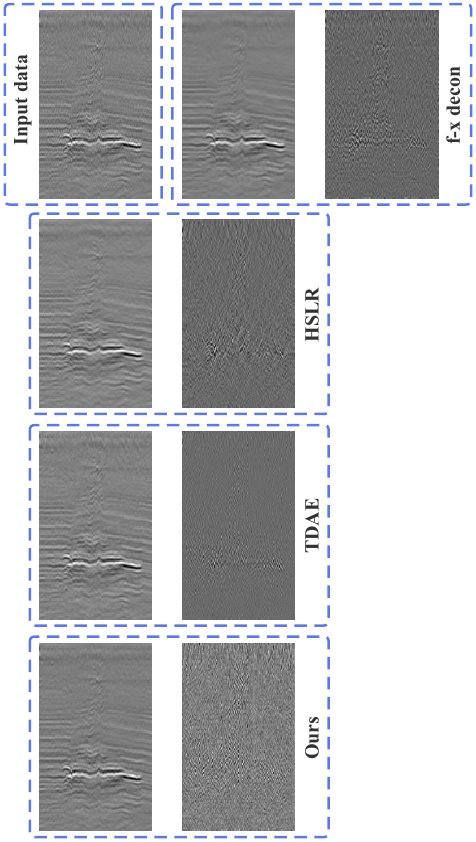}}
\caption{Results and residuals of four methods on $F3\ Netherlands$ patch.}
\end{figure}
\begin{table}[]
\caption{The average local similarity of results on F3 Netherlands (Low is better).}
\centering
\begin{tabular}{|c|c|c|c|}
\hline
f-x decon & HSLR & TDAE & Ours          \\ \hline
0.2008 & 0.1256 & 0.2375 & \textbf{0.1126}         \\ \hline
\end{tabular}
\end{table}

\subsubsection{Experiment on Kerry-3D}

The second data we use comes from $Kerry-3D$, which contains some fault structures and contains some noise. The processing results and residuals of various methods are shown in the Fig 13. Compared with other methods, our proposed method can recognize more noise and also preserve more details. In addition, as in the $F3\ Netherlands$ experiment, we performed the calculation of the average local similarity on the results of various methods. The results are shown in Table 2. In the case of removing a large amount of noise, our proposed method still maintains better signal leakage. In summary, experiments on two sets of field data show that our proposed method can indeed effectively remove noise in seismic data with less signal leakage.
\begin{figure}
\centerline{\includegraphics[width=8.8cm,height=11.253cm]{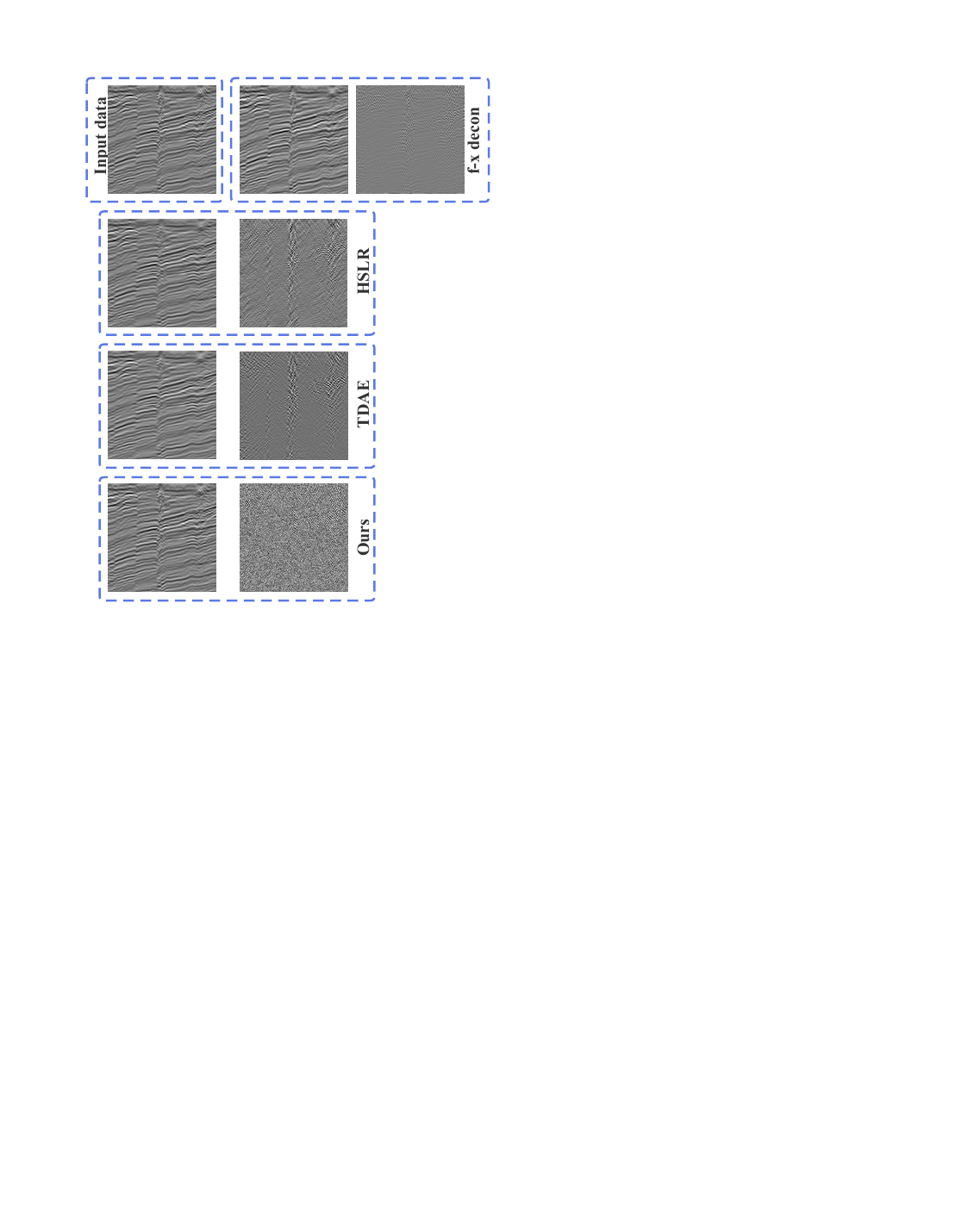}}
\caption{Results and residuals of four methods on $Kerry-3D$ patch.}
\end{figure}
\begin{table}[]
\caption{The average local similarity of results on Kerry-3D (Low is better).}
\centering
\begin{tabular}{|c|c|c|c|}
\hline
f-x decon & HSLR & TDAE & Ours          \\ \hline
0.2969 & 0.2188 & 0.1954 & \textbf{0.1604}         \\ \hline
\end{tabular}
\end{table}

\section{Discussion}

In our present work, we use the diffusion model for noise removal. Diffusion model is a kind of DL model, which was originally proposed in the field of generation. In our work, however, we argue that the process of reconstructing a valid signal can also be viewed as a generative work if it is highly noisy. Using the diffusion model to reverse and gradually restore the signal will also face several problems.

\subsection{Hyperparameters}

In our work, the selection of hyperparameters mainly comes from three aspects, the selection of hyperparameters for DL model, the selection of hyperparameters for diffusion process, and the selection of hyperparameters for noise evaluation.

The DL model involves many parameters, and the relationship between each parameter is complex. After a lot of experiments, we found that as long as the parameters are within a reasonable range, the hyperparameters have little effect on the effect of the model. The model parameters involved in our experiment are as follows:
\begin{equation}
Input=128*128
\end{equation}
\begin{equation}
Downsampling\ Layer = 3
\end{equation}
\begin{equation}
Hidden\ Dimension = [128, 256, 512]
\end{equation}

The hyperparameters in the second part are about the $\beta$ in the diffusion process. The choice of $\beta$ range and step size determines the number of steps for noisy data to be eliminated noise. In our current work, the SNR range of seismic data that we use $\beta$ sequence can be processed is about -2 to 30.

The third part of hyperparameters is about noise evaluation. The parameter involved is mainly the window size. In a large number of our experiments, we found that too small a window may aggravate the saturation phenomenon of the $norm\ factor$, which is not conducive to the evaluation of the noise level; while too large a window may make the $norm\ factor$ concentrate in a small range, and the differentiation is not obvious. After our extensive experiments, a window size of around 9*9 works well. Also, many of the fitted values, such as $A$, depend on the dataset used and have little impact on the estimated performance.

\subsection{The saturation phenomenon of $norm\ factor$}

In our large number of experiments, we found that when the SNR is too small, that is, when the noise content is large, the $norm\ factor$ of the evaluation will tend to a saturated state, and the differentiation under different $t$ conditions is not obvious enough. Therefore, in our future work, we will study how to solve this problem.

\section{Conclusion}

In our present work, we propose the use of diffusion models for denoising seismic data. We use the Bayesian equation to invert the noise addition process, and gradually eliminate the noise in the seismic data to achieve the purpose of restoring the very weak signal. In addition, since the noise removal of seismic data is mostly carried out under the condition of blind Gaussian noise, we propose to use PCA to evaluate the noise, and establish a noise evaluation system to determine the number of steps for reverse noise removal. We conduct experiments on synthetic data and real-world data respectively, and the experiments show that our proposed method can achieve good results in different noise situations, and can identify more noise with less signal leakage. Therefore, we believe that the method we propose has very positive significance for improving the accuracy of seismic exploration and reconstruction of weak signals.

\section*{Acknowledgments}

The code involved in this work has been uploaded to the Github repository: \url{https://github.com/lexiaoheng/DDPM-for-seismic-denoising.git}. Thanks to reviewers and editors for their efforts to improve the quality of our manuscripts.


\bibliography{cite}
\bibliographystyle{IEEEtran}
\end{document}